\documentclass[twocolumn,A4]{article}
\usepackage[dvips]{graphics}
\usepackage{setspace}
\usepackage{color}
\definecolor{gold}{rgb}{0.85,0.66,0}
\definecolor{dblue}{rgb}{0,0,0.5}
\begin{document}
\onecolumn
\begin{center}
{\bf{\Large {\textcolor{gold}{Electron transport through multilevel 
quantum dot}}}}\\
~\\
{\textcolor{dblue}{Santanu K. Maiti}}$^{1,2,*}$ \\
~\\
{\em $^1$Theoretical Condensed Matter Physics Division,
Saha Institute of Nuclear Physics, \\
1/AF, Bidhannagar, Kolkata-700 064, India \\
$^2$Department of Physics, Narasinha Dutt College,
129, Belilious Road, Howrah-711 101, India} \\
~\\
{\bf Abstract}
\end{center}
Quantum transport properties through some multilevel quantum dots sandwiched 
between two metallic contacts are investigated by the use of Green's function 
technique. Here we do parametric calculations, based on the tight-binding 
model, to study the transport properties through such bridge systems. The 
electron transport properties are significantly influenced by (a) number 
of quantized energy levels in the dots, (b) dot-to-electrode coupling 
strength, (c) location of the equilibrium Fermi energy $E_F$ and (d) 
surface disorder. In the limit of weak-coupling, the conductance ($g$) 
shows sharp resonant peaks associated with the quantized energy levels 
in the dots, while, they get substantial broadening in the strong-coupling 
limit. The behavior of the electron transfer through these systems 
becomes much more clearly visible from our study of current-voltage 
($I$-$V$) characteristics. In this context we also describe the noise 
power of current fluctuations ($S$) and determine the Fano factor ($F$) 
which provides an important information about the electron correlation 
among the charge carriers. Finally, we explore a novel transport 
phenomenon by studying the surface disorder effect in which the current 
amplitude increases with the increase of the surface disorder strength 
in the strong disorder regime, while, the amplitude decreases in the 
limit of weak disorder. Such an anomalous behavior is completely 
opposite to that of bulk disordered system where the current amplitude 
always decreases with the disorder strength. It is also observed that 
the current amplitude strongly depends on the system size which reveals
the finite quantum size effect.
\vskip 1cm
\begin{flushleft}
{\bf PACS No.}: 73.23.-b; 73.63.Rt; 85.65.+h \\
~\\
{\bf Keywords}: Green's function; Quantum dot; Conductance; $I$-$V$ 
characteristic; Fano factor; Surface disorder; Bulk disorder.
\end{flushleft}
\vskip 3in
\noindent
{\bf ~$^*$Corresponding Author}: Santanu K. Maiti

Electronic mail: santanu.maiti@saha.ac.in
\newpage
\twocolumn

\section{Introduction}

More advances in nano-science and technology have made feasible to growth
nanometer sized systems, like quantum wires~\cite{til}, quantum 
dots~\cite{hol1,hol2,shan} and molecular wires~\cite{yan}. Electronic 
transport properties in such systems have attracted much more attention 
since these are the fundamental building blocks for future generation of 
electronic devices. There has also been an growing interest in deriving 
analytical results for electron transport in quantum dots, molecular wires 
and single molecule systems. The electron transport properties through 
molecular bridge systems were first studied theoretically in $1974$ by 
Aviram {\em et al.}~\cite{aviram}. Since then several numerous 
experiments~\cite{tali,metz,fish,reed1,reed2} have been performed through 
molecules placed between two electrodes with few nanometer separation. 
Full quantum mechanical treatment is required to characterize the transport 
in such systems. The transport properties are characterized by several 
significant factors like as the quantization of energy levels, quantum 
interference of electron waves~\cite{mag,lau,baer1,baer2,gold,ern1} 
associated with the geometry of the bridging system adopts within the 
junction and other different parameters of the Hamiltonian that are 
needed to describe the complete system. The knowledge of current 
fluctuations (of thermal or quantum origin) also provides several key 
ideas for fabrication of efficient molecular devices. In a review work 
Blanter {\em et al.}~\cite{butt} have described clearly and elaborately 
how the lowest possible noise power of the current fluctuations can be 
determined in a two-terminal conductor. The steady state current 
fluctuations so-called shot noise is a consequence of the quantization 
of charge and it can be used to obtain information on a system which 
is not directly available through conductance measurements. The noise 
power of the current fluctuations gives an additional important 
information about the electron correlation by calculating the Fano factor 
($F$) which directly informs us whether the magnitude of the shot noise 
achieves the Poisson limit ($F=1$) or the sub-Poisson ($F<1$) limit.

Quantum dots are man-made ``droplets" of charge that can contain anything
from a single electron to a collection of several thousand. Their typical 
dimensions range from nanometers to a few microns, and their size, shape 
and interactions can be precisely controlled through the use of advanced
nanofabrication technology. A quantum dot can also be assumed as an 
artificial molecule with few number of atoms and several phenomena can 
be studied by allowing single electron to tunnel into and out of the dot, 
since the quantum dot reveals quantized energy levels, and, here we 
concentrate our study on the electron transport through such a dot. 

There exist several {\em ab initio} methods for the calculation of 
conductance~\cite{yal,ven,xue,tay,der,dam} through a molecular bridge 
system. At the same time the tight-binding model has been extensively 
studied in the literature and it has also been extended to DFT transport
calculations~\cite{elst}.
The study of static density functional theory (DFT)~\cite{kohn} within the
local-density approximation (LDA) to investigate the electron transport
through nanoscale conductors, like atomic-scale point contacts, has met with
great success. But when this similar theory applies to molecular junctions,
theoretical conductances achieve larger values compared to the experimental
predictions and these quantitative discrepancies need extensive study in
this particular field. In a recent work, Sai {\em et al.}~\cite{sai} have
predicted a correction to the conductance using the time-dependent
current-density functional theory since the dynamical effects give
significant contribution in the electron transport, and illustrated some
important results with specific examples. Similar dynamical effects have also
been reported in some other recent papers~\cite{bush,ven1}, where authors have
abandoned the infinite reservoirs, as originally introduced by Landauer, and
considered two large but finite oppositely charged electrodes connected by
a nanojunction. In this article we reproduce an analytic approach based on 
the tight-binding model to characterize the electron transport properties
through some quantum dots placed between two macroscopic contacts. We 
utilize a simple parametric approach~\cite{muj1,san1,muj2,sam,hjo,walc1,
walc2} for these calculations. The model calculations are motivated by 
the fact that the {\em ab initio} theories are computationally much more 
expensive, while, the model calculations by using the tight-binding 
formulation are computationally very cheap and also provide a worth 
insight to the problem. 
In this context we also explore a novel feature of electron transport by 
considering the effect of surface disorder on the dot. Advanced nanoscience 
and technology can easily fabricate a mesoscopic device in which charge 
carriers are scattered mainly by the surface boundaries and not by the 
impurities located in the core region~\cite{ding1,ding2,kou}. The idea 
of such a system named as shell-doped nanowires has been given in recent 
works by Zhong {\em et al.}~\cite{zho1,zho2} where the carrier mobility 
can be controlled
nicely. The shell-doping confines the dopant atoms spatially within a few 
atomic layers in the shell region of nanowire. This is completely opposite 
to that of the conventional doping where the dopant atoms are distributed 
uniformly inside the nanowire. Such a system provides a novel feature of 
electron transport in which the current amplitude increases with the 
increase of the surface disorder strength in the limit of strong disorder, 
while, the current amplitude decreases in the weak disorder limit. To 
emphasize such an interesting phenomenon, here we describe the electron 
transport through a quantum dot in which the impurities are located only 
in its surface boundary. It is also noticed that the electron transport 
through the dot is significantly influenced by the number of quantized
energy levels of the dot which manifests the finite quantum size effects.

Our scheme of study is as follows. Section $2$ describes the theoretical 
formalism of our study. In Section $3$ we focus and explain our significant
results and see that the electron transport properties are significantly
influenced by (a) number of quantized energy levels in the dots, (b)
dot-to-electrode coupling strength, (c) location of $E_F$ and (d)
surface impurity. Finally, we draw our conclusions in Section $4$.

\section{Description of model and formalism}

\begin{figure}[ht]
{\centering \resizebox*{6.75cm}{3.25cm}{\includegraphics{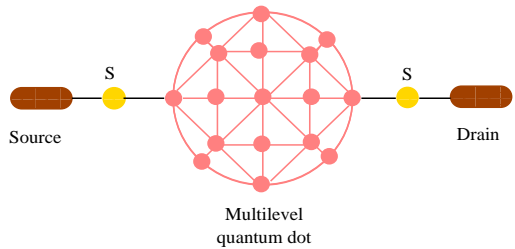}}\par}
\caption{(Color online). Schematic representation of a multilevel quantum 
dot attached to two metallic electrodes (source and drain) through lattice 
sites $S$ and $S$ where the filled red circles correspond to the atomic 
sites in the dot.}
\label{quantumdot}
\end{figure}

Here we describe very briefly the methodology for the calculation of the 
transmission probability ($T$), conductance ($g$), current ($I$) and the 
noise power of its fluctuations ($S$) through a quantum dot attached to two 
metallic electrodes (schematically illustrated as in Fig.~\ref{quantumdot}) 
by using the Green's function technique.

At sufficient low temperature and small applied voltage the conductance $g$ 
of the dot is expressed through the Landauer conductance formula~\cite{datta},
\begin{equation}
g=\frac{2e^2}{h} T
\label{equ1}
\end{equation}
where the transmission probability $T$ is written in this form~\cite{datta},
\begin{equation}
T={\mbox{Tr}}\left[\Gamma_S G_{dot}^r \Gamma_D G_{dot}^a\right]
\label{equ2}
\end{equation}
In this expression $G_{dot}^r$ and $G_{dot}^a$ are the retarded and advanced 
Green's functions of the dot and $\Gamma_S$ and $\Gamma_D$ describe its (dot) 
coupling to the source and drain, respectively. The Green's function of 
the multilevel quantum dot is expressed as,
\begin{equation}
G_{{\mbox{dot}}}=\left(E-H_{{\mbox{dot}}}-\Sigma_S-\Sigma_D\right)^{-1}
\label{equ3}
\end{equation}
where $E$ is the energy of the injecting electron and $H_{dot}$ is the 
Hamiltonian of the dot (here the quantum dot is assumed as an artificial
molecule with several number of atoms) which can be written in the 
tight-binding model within the non-interacting picture like, 
\begin{equation}
H_{{\mbox{dot}}}=\sum_i \epsilon_i c_i^{\dagger} c_i + \sum_{<ij>} t 
\left(c_i^{\dagger} c_j + c_j^{\dagger} c_i\right)
\label{equ4}
\end{equation}
where $\epsilon_i$'s are the site energies and $t$ is the hopping strength
between two nearest-neighbor atomic sites in the dot. To introduce the 
impurities in the dot we choose the site energies ($\epsilon_i$'s) in the 
form of incommensurate potentials through the expression $\epsilon_i=\sum_i W 
\cos(i \lambda \pi)$ where $\lambda$ is an irrational number and $W$ is the
strength of the disorder. As a typical example we take the value of $\lambda$ 
as the golden mean $\left(1+\sqrt{5}\right)/2$. Setting $\lambda=0$ we get 
back the pure system with identical site potential $W$. In Eq.(\ref{equ3}), 
$\Sigma_S$ and $\Sigma_D$ correspond to the self-energies due to coupling 
of the dot to the two electrodes. These two semi-infinite one-dimensional 
metallic electrodes are described also by the similar kind of tight-binding 
Hamiltonian as given in Eq.(\ref{equ4}), where we take the site energy and 
the nearest-neighbor hopping strength by the parameters $\epsilon_i^{\prime}$ 
and $v$, respectively. All the information about the dot-to-electrode 
coupling are included into these two self-energies as stated above and 
are described through the use of Newns-Anderson chemisorption 
theory~\cite{muj1,muj2}. The detailed description of this theory is 
available in these two references. By utilizing the Newns-Anderson type 
model we can express the conductance in terms of the effective dot 
properties multiplied by the effective state densities involving the 
coupling. This allows us to study directly the conductance as a function 
of the properties of the electronic structure of the dot between the 
electrodes.

The current passing through the dot can be considered as a single electron
scattering process between the two reservoirs of charge carriers. The
current-voltage ($I$-$V$) relationship can be computed from the 
expression~\cite{datta},
\begin{equation}
I(V)=\frac{e}{\pi \hbar}\int \limits_{-\infty}^{\infty} 
\left(f_S-f_D\right) T(E) dE
\label{equ5}
\end{equation}
where $f_{S(D)}=f\left(E-\mu_{S(D)}\right)$ gives the Fermi distribution
function with the electrochemical potentials $\mu_{S(D)}=E_F\pm eV/2$. For 
the sake of simplicity, here we assume that the entire voltage is dropped 
across the dot-electrode interfaces and this assumption does not greatly 
affect the qualitative aspects of the $I$-$V$ characteristics. Such an 
assumption is based on the fact that the electric field inside the dot, 
especially for the dots with smaller number of atomic sites, seems to 
have a minimal effect on the conductance-voltage characteristics. On the 
other hand, for the dots with very large number of atomic sites and high 
bias voltage, the electric field inside the dot may play a more 
significant role depending on the internal structure of the dot~\cite{tian}, 
yet the effect is much small.

The noise power of the current fluctuations is calculated through the 
following relation~\cite{butt},
\begin{eqnarray}
S & = & \frac{2e^2}{\pi \hbar}\int \limits_{-\infty}^{\infty}\left[T(E)
\left\{f_S\left(1-f_S\right) + f_D\left(1-f_D\right) \right\} \right. 
\nonumber \\
 & & + T(E) \left. \left\{1-T(E)\right\}\left(f_S-f_D\right)^2 \right] dE
\label{equ6}
\end{eqnarray}
where the first two terms in this equation correspond to the equilibrium
noise contribution and the last term gives the non-equilibrium or shot noise
contribution to the power spectrum. By calculating the noise power of the
current fluctuations we can evaluate the Fano factor $F$, which is essential
to predict whether the shot noise lies in the Poisson or the sub-Poisson
regime, through the relation~\cite{butt},
\begin{equation}
F=\frac{S}{2 e I}
\label{equ7}
\end{equation}
The shot noise achieves the Poisson limit when $F=1$ and in this case no 
electron correlation exists between the charge carriers. On the other hand,
for $F<1$ the shot noise reaches the sub-Poisson limit and it provides 
the information about the electron correlation among the charge carriers.

In this article we perform all the calculations at absolute zero temperature, 
but the qualitative behavior of all the results are invariant up to some 
finite (low) temperature. The reason for such an assumption is that the 
broadening of the energy levels of the dot due to its coupling to the 
electrodes is much larger than that of the thermal broadening. For 
simplicity, we take the unit $c=e=h=1$ in our present investigation.

\section{Results and discussion}

This section demonstrates the transport properties of some multilevel
\begin{figure}[ht]
{\centering \resizebox*{8cm}{11cm}{\includegraphics{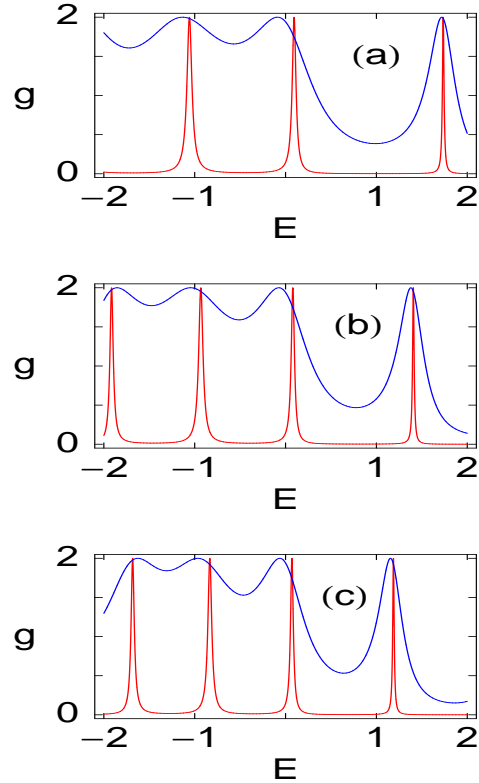}}\par}
\caption{(Color online). $g$-$E$ characteristics for the multilevel 
quantum dots, where (a), (b) and (c) are respectively for the dots 
with $14$, $18$ and $22$ atomic sites. The red and blue curves 
correspond to the weak- and strong-coupling cases, respectively.}
\label{dotcond}
\end{figure}
quantum dots in the coherent transport regime. In the bridge system, the 
dot is attached to two metallic electrodes through the lattice sites $S$ 
and $S$, as schematically illustrated in Fig.~\ref{quantumdot}. In actual
experimental set up, these two electrodes made from gold are used and the
dot attached to them via thiol groups in the chemisorption technique and 
in making such contact with these electrodes, the hydrogen (H) atoms of the 
thiol groups remove and the sulfur (S) atoms reside. In our tight-binding
formulation, the quantum dot is coupled to the electrodes through the 
lattice sites $S$ and $S$ by using the parameters $\tau_S$ and $\tau_D$, 
where they (coupling parameters) correspond to the coupling strengths to 
the source and drain, respectively. Here we concentrate our results on 
clarifying the following points: the dependence of the conductance, current
and the noise power of its fluctuations on (I) the number of energy levels 
in the dot and (II) dot-to-electrode coupling strength and also discuss 
the dependence of the current and its fluctuations on the location of the 
equilibrium Fermi energy $E_F$. Finally, attention is drawn on the study
of the surface disorder effect in such electron transport. Throughout this 
article we will discuss all the essential features of the electron transport 
for the two distinct regimes. One is the so-called weak-coupling regime 
denoted by the condition $\tau_{\{S,D\}} << t$ and the other one is the
so-called strong-coupling regime where $\tau_{\{S,D\}} \sim t$. The values 
of such parameters for these two distinct regimes are chosen as 
$\tau_S=\tau_D=0.5$, $t=2.5$ (weak-coupling) and $\tau_S=\tau_D=2$, $t=2.5$ 
(strong-coupling).

The characteristic behavior of the conductance $g$ as a function of the
injecting electron energy $E$ for the multilevel quantum dots are shown
in Fig.~\ref{dotcond}, where (a), (b) and (c) correspond to the dots with 
$14$, $18$ and $22$ atomic sites, respectively. The red lines represent 
the results in the limit of weak-coupling, while, the blue lines denote 
the results for the strong-coupling limit. In the weak-coupling limit, 
the conductance shows very sharp resonant peaks (red curves in 
Fig.~\ref{dotcond}) for some particular energy values, while, for all other 
energies it almost vanishes. At these resonances the conductance $g$ achieves 
the value $2$, and accordingly, the transmission probability $T$ goes to 
unity, since from the Landauer conductance formula we get $g=2T$ (see 
Eq.(\ref{equ1}) with $e=h=1$ in our present description). These resonant 
peaks are associated with the energy eigenvalues of the corresponding dot 
and therefore more resonant peaks appear with the increase of the quantized 
energy levels in the dot. Thus it can be emphasized that the conductance 
spectrum manifests itself the energy eigenvalues of the dot. With the 
increase of the dot-to-electrode coupling strength, the widths of these 
resonances get enhanced substantially, as illustrated by the blue curves in 
\begin{figure*}[ht]
{\centering \resizebox*{4.5cm}{11cm}{\includegraphics{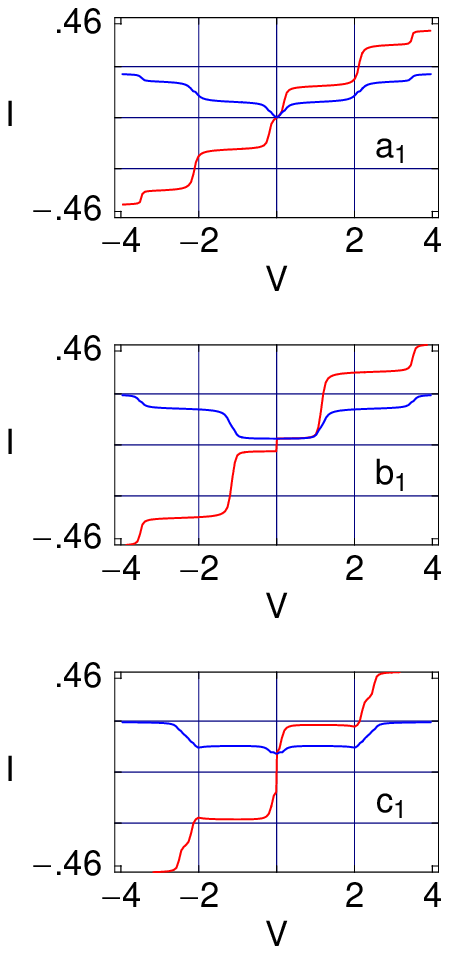}}
\resizebox*{4.5cm}{11cm}{\includegraphics{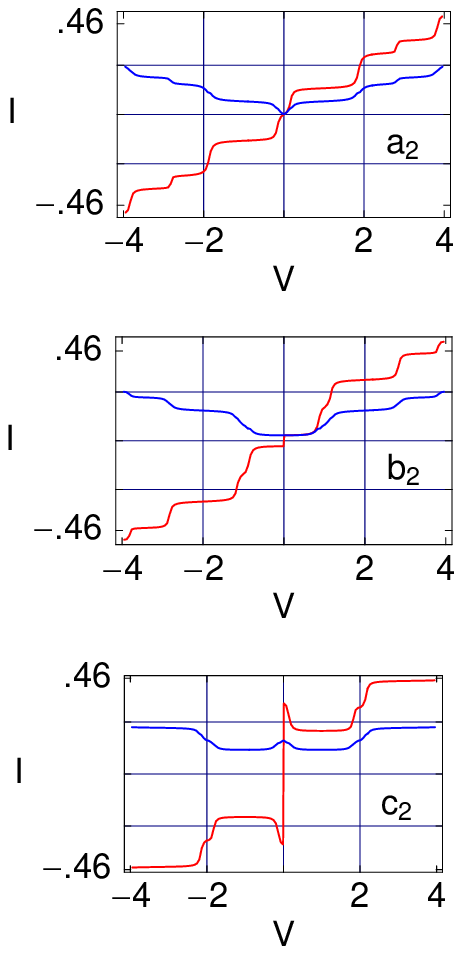}}
\resizebox*{4.5cm}{11cm}{\includegraphics{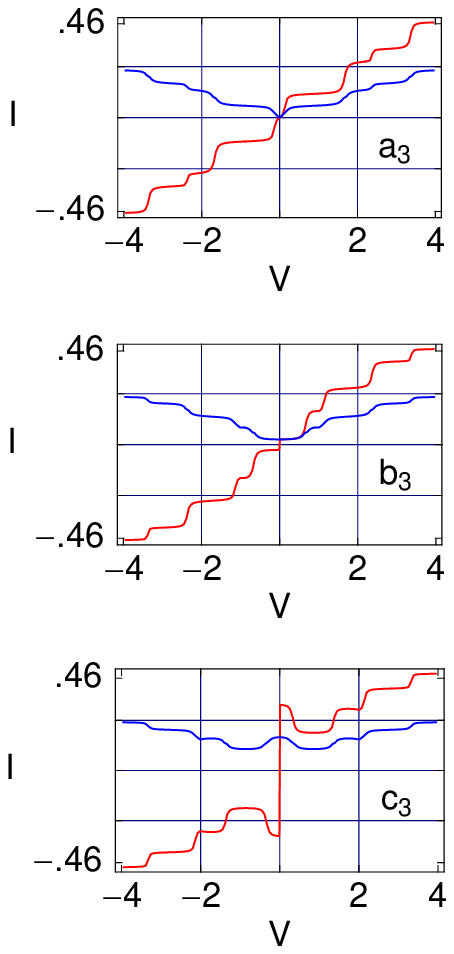}}\par}
\caption{(Color online). Current $I$ (red curve) and the noise power of 
its fluctuations $S$ (blue curve) as a function of the applied bias 
voltage $V$ for the multilevel dots in the limit of weak-coupling, where 
the $1$st, $2$nd and $3$rd columns are respectively for the dots with 
$14$, $18$ and $22$ atomic sites. The $1$st, $2$nd and $3$rd rows 
correspond to the results for the dots with the Fermi energy $E_F=0$, 
$-0.5$ and $-1$, respectively.}
\label{dotcurrlow}
\end{figure*}
Fig.~\ref{dotcond}. This is due to the substantial broadening of the 
quantized energy levels in the limit of strong-coupling. The contribution 
for such broadening of the energy levels comes from the imaginary parts 
of the two self energies $\Sigma_S$ and $\Sigma_D$, 
respectively~\cite{datta}. Thus for the strong-coupling limit, the electron 
conducts through the dots for the wide range of energies, while, a fine 
tuning in the energy scale is necessary to get the electron conduction 
through these systems in the limit of weak-coupling. Therefore, it can 
be predicted that the dot-to-electrode coupling strength has a significant 
role in the
\begin{figure*}[ht]
{\centering \resizebox*{4.5cm}{11cm}{\includegraphics{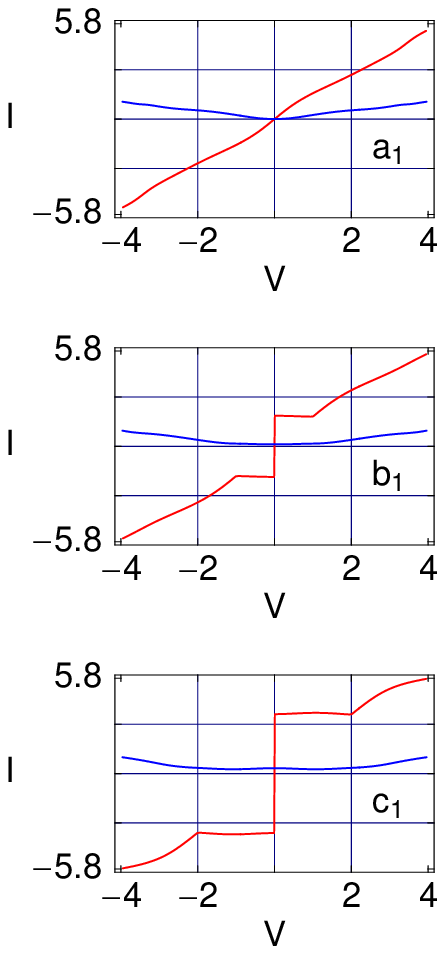}}
\resizebox*{4.5cm}{11cm}{\includegraphics{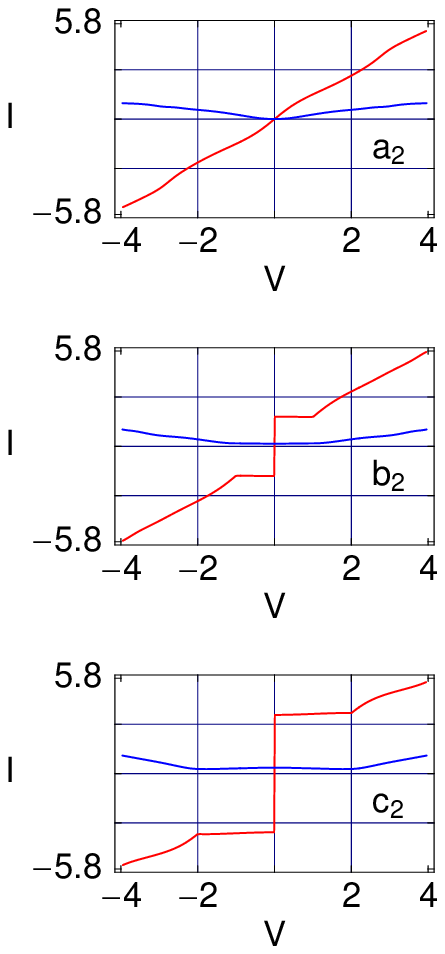}}
\resizebox*{4.5cm}{11cm}{\includegraphics{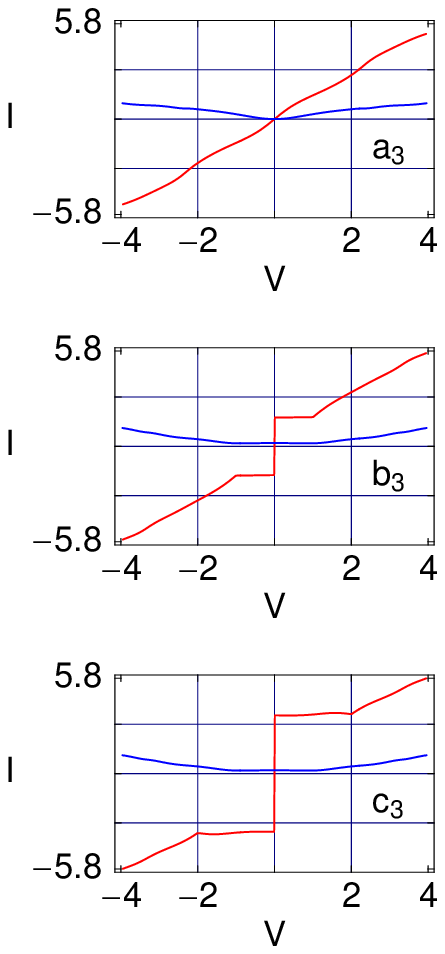}}\par}
\caption{(Color online). Current $I$ (red curve) and the noise power of 
its fluctuations $S$ (blue curve) as a function of the applied bias 
voltage $V$ for the multilevel dots in the limit of strong-coupling, 
where the $1$st, $2$nd and $3$rd columns are respectively for the dots 
with $14$, $18$ and $22$ atomic sites. The $1$st, $2$nd and $3$rd rows 
correspond to the results for the dots with the Fermi energy $E_F=0$, 
$-0.5$ and $-1$, respectively.}
\label{dotcurrhigh}
\end{figure*}
determination of the electron conduction through the bridge systems. This 
feature provides a key information in the study of molecular transport 
phenomena.

The behavior of the electron transfer through such systems can be described 
much more clearly by studying the current-voltage ($I$-$V$) characteristics.
In the forthcoming parts we will concentrate our study on the current and 
the noise power of its fluctuations ($S$) as a function of the applied
bias voltage ($V$) for these quantum dots. Both the current and the noise 
power of its fluctuations are determined from the integration method of the
transmission function ($T$) as described in Eq.(\ref{equ5}) and in
Eq.(\ref{equ6}), where the transmission function varies exactly similar
to that of the conductance spectra as illustrated in Fig.~\ref{dotcond},
differ only in magnitude by the factor $2$ since the relation $g=2T$ holds
from the Landauer conductance formula (Eq.(\ref{equ1})).
In Fig.~\ref{dotcurrlow}, we display the variation of the current and the
noise power of its fluctuations as a function of the applied bias voltage 
for the multilevel quantum dots in the limit of weak-coupling, where the
first, second and third columns are respectively for the dots with
$14$, $18$ and $22$ atomic sites. The red line corresponds to the current 
and the blue line represents the noise power of its fluctuations. In order 
to emphasize the effect of the location of the Fermi energy $E_F$ on such 
transport here we plot the results considering three different values of 
$E_F$, where the first, second and third rows correspond to the results 
for the dots with $E_F=0$, $-0.5$ and $-1$, respectively. The current shows 
staircase-like structure with sharp steps as a function of the applied bias 
voltage. This is due to the existence of the sharp resonant peaks in the 
conductance spectra (see the red curves in Fig.~\ref{dotcond}) in this 
limit of weak-coupling, since we compute the current from the integration 
procedure of the transmission function $T$. The electrochemical potentials 
on the electrodes are shifted gradually with the increase of the applied 
bias voltage and eventually cross one of the quantized energy levels in 
the dot. Accordingly, a current channel is opened up and the current-voltage 
characteristic curve produces a jump. With the 
increase of the dot size i.e., number of quantized energy levels, current 
shows more steps (as expected) which is clearly visible from this 
Fig.~\ref{dotcurrlow}. For all these bridges we observe that the current 
amplitudes are too small and they are comparable with each other (see the 
red lines in Fig.~\ref{dotcurrlow}). Now we discuss the effect of the 
location of the Fermi energy $E_F$ in these quantum dots. The effect is 
quite interesting. We see that for $E_F=0$, the current amplitude for all 
these three bridges across $V=0$ is almost zero, while, the amplitude 
gradually increases across this voltage ($V=0$) as we change the Fermi 
energy $E_F$ to the values $-0.5$ and $-1$, respectively.
For $E_F=-1$, we get a very large current compared to the other two values
of $E_F$. For a particular coupling strength this current amplitude depends
on both the location of the Fermi energy and the number of the quantized 
energy levels in the dots. Thus it can be predicted that, by tuning the 
Fermi energy we can get the on/off state of the bridge system across $V=0$. 
This is an important finding in the study of molecular transport. Now in the 
determination of the noise power of the current fluctuations we get several 
interesting results depending on the values of $E_F$ and the number of 
quantized energy levels. Both for the choices of $E_F=0$ and $-1$, the shot 
noise (blue curves in the $1$st and the $3$rd rows of Fig.~\ref{dotcurrlow}) 
lies in the sub-Poisson regime ($F<1$) momentarily as we switch on the bias 
voltage. Accordingly, for such cases the electrons are always correlated 
with each other. Here the correlation of the electrons means one electron 
feels the existence of the other in the sense of Pauli 
exclusion principle, since no other electron-electron interaction is taken
into account in our present description. On the other hand for the case
where we set $E_F=-0.5$, the shot noise (blue curves in the $2$nd row of
Fig.~\ref{dotcurrlow}) makes a transition from the Poisson limit ($F=1$)
to the sub-Poisson limit ($F<1$) as long as we cross the first step in
the current-voltage characteristics. This indicates that the electrons are 
correlated after the tunneling process has occurred. For such a particular
case ($E_F=-0.5$), it is also observed that the threshold bias voltage 
($V_{th}$) where the shot noise makes a transition from the Poisson to
the sub-Poisson limit gradually decreases with the increase of the quantized
energy levels in the quantum dots. Another important observation is that,
for all these three bridge systems the noise power of the current fluctuations 
remains in the same level independent of the number of the quantized energy
levels i.e., the number of atomic sites in the dots.

The characteristic features of the current and the noise power of its 
fluctuations are also very interesting for these bridges in the 
limit of strong-coupling. The results are shown in Fig.~\ref{dotcurrhigh},
where the figures in the different rows and columns correspond to the 
same meaning as presented in Fig.~\ref{dotcurrlow}. The red and blue
curves also represent the identical meaning as in Fig.~\ref{dotcurrlow}.
From the results plotted in Fig.~\ref{dotcurrhigh} it is observed that,
both the current and noise power vary quite continuously as a function
of the applied bias voltage $V$. Such kind of behavior appears due to the
broadening of the resonant peaks (blue curves in Fig.~\ref{dotcond}) in 
the limit of strong-coupling, since the current and the noise power are
determined from the integration procedure of the transmission function $T$. 
One key result is that, the current amplitudes get enhanced quite 
significantly compared to the current amplitudes obtained in the limit 
of weak-coupling (see the red curves in Fig.~\ref{dotcurrlow}). This can 
be understood by noting the areas under the curves in the conductance 
spectra for the two limiting cases as plotted in Fig.~\ref{dotcond}. 
Thus by tuning the dot-to-electrode coupling strength one can achieve 
greater current across the bridge system which also provides an 
interesting phenomenon in fabrication of molecular devices. Lastly, in 
the study of the noise power of the current fluctuations we find that, 
for all the choices of the Fermi energy $E_F$
there is no such possibility of transition from the Poisson limit to the
sub-Poisson limit, since the shot noise already achieves the sub-Poisson
limit (blue curves of Fig.~\ref{dotcurrhigh}) momentarily as we apply
the bias voltage. Therefore, for all such values of $E_F$ the electron 
correlation is more significant. Thus we can emphasize that, both the
dot-to-electrode coupling strength and the location of the Fermi energy
are the key factors that control the electron transport through a bridge
system.

\begin{figure}[ht]
{\centering \resizebox*{6.75cm}{3.25cm}{\includegraphics{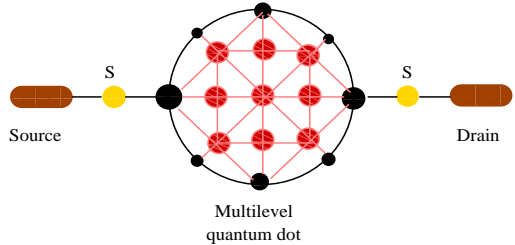}}\par}
\caption{(Color online). Schematic view of a surface disordered multilevel 
quantum dot attached to two metallic electrodes (source and drain) through 
lattice sites $S$ and $S$. Disorder in the surface of the dot is represented 
by the filled black circles of different sizes which correspond to the 
different lattice sites with variable site energies.}
\label{dotbridge}
\end{figure}

\begin{figure}[ht]
{\centering \resizebox*{8cm}{5.5cm}{\includegraphics{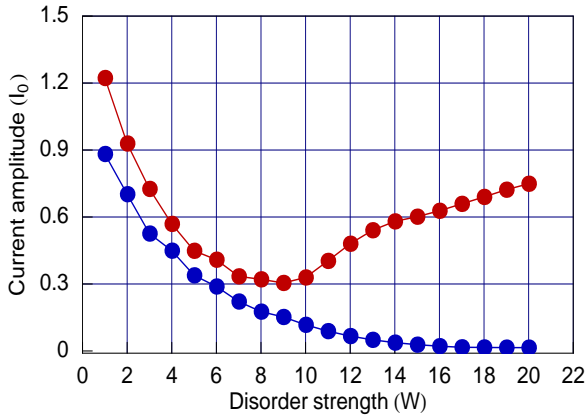}}\par}
\caption{(Color online). Current amplitudes ($I_0$) as a function of the 
disorder strength ($W$) for the multilevel quantum dots with $35$ atomic 
sites in the strong-coupling limit where we take $N_s=20$ and $N_c=15$. 
The red and blue curves correspond to the surface and bulk disordered 
cases, respectively. The typical current amplitudes are computed for 
$E_F=0$.}
\label{surface1}
\end{figure}

Now we concentrate our study on the correlation effect between the surface
disorder and bulk disorder on the electron transport through the multilevel 
quantum dots. The schematic representation of a surface disordered 
multilevel quantum dot attached to two metallic electrodes is shown in 
Fig.~\ref{dotbridge}. The disorder in the surface is represented by the 
filled black circles of different sizes which correspond to the different 
atomic sites with variable site energies, while, the inner core perfect 
region is described by the filled red circles with identical site energies. 
In order to introduce the impurities, we choose the site energies 
($\epsilon_i$'s) from the incommensurate potential distribution function 
as stated earlier in Section $2$. For the surface disordered system, the 
impurities are
given only in the atomic sites located in the surface, while, for the bulk
disordered case the impurities are introduced in all the atomic sites. Here 
we use the parameters $N_s$ and $N_c$ to denote the total number of atomic 
sites in the surface boundary and in the core region of the dot, respectively.
For the sake of simplicity, we set the equilibrium Fermi energy $E_F=0$ in 
this particular study. Figure~\ref{surface1} shows the variation of the 
current amplitudes ($I_0$), in the strong-coupling limit, as a function of 
the impurity strength ($W$) for the multilevel quantum dots with $N_s=20$ 
and $N_c=15$. The current amplitudes are computed at the typical bias 
voltage $V=1.5$, where the red and blue curves correspond to the results 
for the surface and bulk disordered cases, respectively. Since we introduce 
only the diagonal disorder by 
considering the site energies from the known incommensurate potential 
distribution function ($\epsilon_i=\sum_i W\cos(i \lambda \pi)$ with $\lambda=
\left(1+\sqrt{5}\right)/2$) we do not take any disorder averaging during our 
calculation. The idea of considering such kind of potential distribution
function rather than any random distribution function is to avoid the disorder 
averaging over large number of possible disordered configurations since it 
takes too much time to evaluate the results. Now instead of considering such 
an
unconventional incommensurate potential function, we can also take any random 
distribution function and in that case we have to take the average over a 
large number of random disordered configurations to achieve much more accurate 
result. Both these two different treatments of the disorder in the model
provide quite similar in nature for the variation of the current amplitude 
and due to this fact we choose the unconventional treatment of the disorder,
instead of the other one, to understand the results through limited numerical 
resources. From the results
it is observed that, in the bulk disordered case the current amplitude
gradually decreases with the increase of the impurity strength and for the
strong enough impurity it almost drops to zero. This behavior can be well
understood from the theory of Anderson localization where the states become 
more localized with the increase of the impurity strength. The significant 
feature appears when the impurities are given only in the surface boundary 
of the dot. The current amplitude initially decreases with the strength of
the impurity, while, beyond some critical value of the impurity strength
$W=W_c$ (say) the amplitude increases. Such an anomalous behavior is 
completely opposite to that of the bulk disordered case and it can be 
explained as follows. In the ordered-disordered separated quantum dot, a
gradual separation of the energy spectra of the disordered surface and the
perfect inner core regions takes place with the increase of the disorder
strength $W$. Accordingly, the influence of random scattering in the 
perfect region due to the strong localization in the disordered surface
region decreases. This can be mathematically implemented in such a way.
For an ordered-disordered separated quantum dot, we can write the effective
Hamiltonian for the disordered surface as $H_S^{*}=H_S-\xi(W)$, while for
the inner core perfect region the effective Hamiltonian becomes
$H_C^{*}=H_C-\eta(W)$. Here $\xi(W)=H_{SC}\left(H_C-E\right)^{-1}H_{CS}$
and $\eta(W)=H_{CS}\left(H_S-E\right)^{-1}H_{SC}$, with $H_C$ and $H_S$
are the original sub-Hamiltonians for the perfect inner core and the 
disordered surface regions, respectively, and $H_{SC} (H_{CS})$ describes
their interaction. For $|E|<<W$, we get $\eta \sim H_{CS} H_S^{-1} H_{SC}$,
leading to $\eta(W)\rightarrow 0$ as $W\rightarrow \infty$. This reveals
that, the energy spectrum of an ordered-disordered separated quantum dot
with large disorder contains localized tail states with much small and
central states with much large values of localization length, contributed 
approximately by $H_S$ and $H_C$, respectively. Thus the central states 
gradually separated from the tail states and delocalized with the increase
of the strength of the disorder. Thus we see that, for the coupled
ordered-disordered separated quantum dot system, the coupling between the
localized states with the inner core extended states is strongly 
influenced by the strength of the surface disorder, and, this coupling is
inversely proportional to the disorder strength $W$. Therefore, in the weak 
disorder regime the coupling effect is strong, while, the coupling effect 
becomes less significant in the limit of strong disorder. Accordingly, in 
the weak disorder
regime the electron transport is strongly influenced by the impurities at
the surface such that the electron states are scattered more and hence
the current amplitude decreases. On the other hand, for the stronger disorder
regime the inner core extended states are less influenced by the surface 
disorder and the coupling effect gradually decreases with the increase of 
the impurity strength which provide the larger current amplitude in the 
strong disorder regime. For large enough impurity strength, the inner core 
extended states are almost unaffected by the impurities at the surface 
boundary and in that case the current is carried only by these inner core 
extended states which is the trivial limit. So the exciting limit is the 
intermediate limit of $W$.

\begin{figure}[ht]
{\centering \resizebox*{8cm}{5.5cm}{\includegraphics{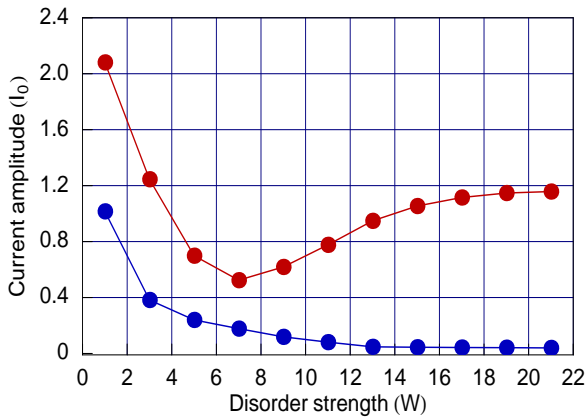}}\par}
\caption{(Color online). Current amplitude ($I_0$) as a function of 
the disorder strength ($W$) for the multilevel quantum dots with $49$ 
atomic sites in the strong-coupling limit where we consider $N_s=24$ 
and $N_c=25$. The red and blue curves represent the identical meaning 
as in Fig.~\ref{surface1}. The typical current amplitudes are computed 
for $E_F=0$.}
\label{surface2}
\end{figure}

To reveal the finite quantum size effects on the electron transport now we 
focus our results for the other system size where we consider $N_s=24$ and 
$N_c=25$. The results are plotted in Fig.~\ref{surface2}, in the limit of 
strong dot-to-electrode coupling, where all the current amplitudes are 
computed at the typical bias voltage $V=1.5$ (same as earlier). The red 
and blue lines denote the identical meaning as in Fig.~\ref{surface1}. 
Both for the surface and bulk disordered systems the current amplitudes 
show almost the similar behavior for the two different disordered regimes 
as predicted in Fig.~\ref{surface1}. But the significant observation is 
that the overall current amplitude for this bridge ($N_s=24$ and $N_c=25$) 
in the case of surface disorder is much larger compared to the results 
as observed
previously i.e., for the surface disordered quantum dot with $N_s=20$ and 
$N_c=15$ (see the red line of Fig.~\ref{surface1}). This behavior can be 
explained in such a way. The ratio of the atomic sites in the surface region 
to the atomic sites in the inner core region for the quantum dot with $49$ 
atomic sites is much smaller than that of the dot with $35$ atomic sites and 
accordingly, the surface effect becomes much less significant for the dot 
with $49$ atomic sites compared to the other dot. Therefore, the current 
carried by the states in the inner core region for this dot will be less 
affected by the surface disorder which provides greater current amplitude. 
Another important observation is that, the typical current amplitude where 
it goes to a minimum strongly depends on the system size i.e., the number 
of quantized energy levels or the total number of atomic sites in the dot 
which is clearly visible from the red curves illustrated in 
Figs.~\ref{surface1} and \ref{surface2}. These results reveal the finite 
quantum size effects in the study of the electron transport phenomena. The
underlying physics behind the location of the minimum in the current versus
disorder curve is quite interesting. Two competing mechanisms are there that
control the current amplitude. One is the random scattering in the inner
core perfect region due to the localization in the disordered surface which
tends to decrease the current, and, the other one is the vanishing influence
of random scattering in the ordered region due to the strong localization 
in the disordered surface which provides the enhancement of the current.
Now, depending on the ratio of the atomic sites in the surface region to the
atomic sites in the inner core region, the vanishing effect of random 
scattering from the ordered states dominates over the non-vanishing effect 
of random scattering from these states for a particular disorder strength 
$(W=W_c)$ which provides the location of the minimum in the current versus 
disorder curve.  

A Similar feature of the surface disorder effect is also observed in the 
limit of weak dot-to-electrode coupling strength with reduced current 
amplitudes for these quantum dots and in the obvious reason here we do 
not describe these results once again. Throughout our study of the 
surface disorder effect on the electron transport we compute all the 
typical current amplitudes for the equilibrium Fermi energy $E_F=0$ 
only, and, this peculiar behavior will also be observed for the other 
values of $E_F$.

\section{Concluding remarks}

In conclusion of this article, we have introduced a parametric approach 
based on the tight-binding model to study the electron transport 
characteristics through some multilevel quantum dots. From our results 
we can predict that the electron transport is significantly influenced by
(a) the number of quantized energy levels in the dots, (b) the location
of the equilibrium Fermi energy $E_F$, (c) the dot-to-electrodes coupling 
strength and (d) the surface disorder. All the results have been performed 
by using the Green's function technique and this technique can be used to 
study the electron transport in any complicated system, like complicated 
organic molecule, quantum wire, array of quantum dots etc., which bridges 
the two reservoirs.

The conductance shows sharp resonant peaks for the weak-coupling limit
(red curves of Fig.~\ref{dotcond}), while, they get broadened in the limit 
of strong-coupling (blue curves of Fig.~\ref{dotcond}). Such increment of 
the resonant widths is due to the broadening of the quantized energy levels
of the dots, where the contribution comes from the imaginary parts of the 
two self energies $\Sigma_S$ and $\Sigma_D$~\cite{datta}.

In the determination of the current, we have seen that the current shows
staircase-like structures with sharp steps (red lines in Fig.~\ref{dotcurrlow})
in the limit of weak-coupling, while it (current) varies quite continuously 
(red lines in Fig.~\ref{dotcurrhigh}) and achieves very large value in the 
strong-coupling limit. 

Next in the description of the noise power of the current fluctuations we have 
noticed that whether the shot noise lies in the Poisson regime ($F=1$) or in 
the sub-Poisson regime ($F<1$) strongly depends on the location of the Fermi 
energy $E_F$ and the dot-to-electrodes coupling strength.

Finally, in the study of the surface disorder effect we have explored a novel 
transport phenomenon in which the current amplitude increases with the 
increase of the surface disorder strength in the strong disorder regime, 
while, the current amplitudes decreases in the weak disorder regime. Such an 
anomalous behavior has not been pointed out previously in the literature where 
the transport properties have been described through the bridge systems. This 
feature is completely opposite to that of the bulk disordered system in which 
the current amplitude decays gradually with the increase of the impurity 
strength and eventually drops to zero.

Several realistic assumptions have been made in the present study. More 
studies are expected to take into account the Schottky effect which comes 
from the charge transfer across the dot-electrode interfaces, the static 
Stark effect, which is considered for the modification of the electronic 
structure of the bridging system due to the applied bias voltage (essential 
especially for higher voltages). However all these effects can be included 
into our framework by a simple generalization of the presented formalism. 
In this article we have also neglected the effects of all the inelastic 
scattering processes and the Coulomb correlation to characterize the electron 
transport through such quantum dots.

\end{document}